\documentclass[reprint,aip,apl,superscriptaddress]{revtex4-2}

\usepackage{graphicx}
\usepackage{dcolumn}
\usepackage{amsmath}
\usepackage{amssymb}
\usepackage{float}
\usepackage{pifont}
\usepackage{textgreek}
\usepackage{textcomp}
\usepackage{gensymb}
\usepackage{rotating}
\usepackage[hidelinks]{hyperref}

\hypersetup{
  colorlinks   = true,
  urlcolor   = blue,
  citecolor   = blue
}

\begin{document}

\title{Defect engineering-induced Seebeck coefficient and carrier concentration decoupling in CuI by noble gas ion implantation}

\author{Martin Markwitz}\email{martin.markwitz@vuw.ac.nz}
 \affiliation{School of Chemical and Physical Sciences, Victoria University of Wellington, PO Box 600, Wellington 6140, New Zealand}
 \affiliation{National Isotope Centre, GNS Science, PO Box 30368, Lower Hutt 5010, New Zealand}
 \affiliation{The MacDiarmid Institute for Advanced Materials and Nanotechnology, Victoria University of Wellington, PO Box 600, Wellington 6140, New Zealand}
\author{Peter P. Murmu}
 \affiliation{National Isotope Centre, GNS Science, PO Box 30368, Lower Hutt 5010, New Zealand}
\author{Takao Mori}
 \affiliation{International Center for Materials Nanoarchitectonics (WPI-MANA), National Institute for Materials Science (NIMS), 1-1 Namiki, Tsukuba, Ibaraki 305-0044, Japan}
 \affiliation{Graduate School of Pure and Applied Science, University of Tsukuba, 1-1-1 Tennodai, Tsukuba, Ibaraki 305–8671, Japan}
\author{John V. Kennedy}
 \affiliation{National Isotope Centre, GNS Science, PO Box 30368, Lower Hutt 5010, New Zealand}
 \affiliation{The MacDiarmid Institute for Advanced Materials and Nanotechnology, Victoria University of Wellington, PO Box 600, Wellington 6140, New Zealand}
\author{Ben J. Ruck}
 \affiliation{School of Chemical and Physical Sciences, Victoria University of Wellington, PO Box 600, Wellington 6140, New Zealand}
 \affiliation{The MacDiarmid Institute for Advanced Materials and Nanotechnology, Victoria University of Wellington, PO Box 600, Wellington 6140, New Zealand}

\date{\today}

\begin{abstract}
Copper(I) iodide, CuI, is the leading $p$-type non-toxic and earth-abundant semiconducting material for transparent electronics and thermoelectric generators. Defects play a crucial role in determining the carrier concentration, scattering process, and therefore thermoelectric performance of a material. A result of defect engineering, the power factor of thin film CuI was increased from $332\pm32$\,{\textmu}Wm\textsuperscript{-1}K\textsuperscript{-2} to $578\pm58$\,{\textmu}Wm\textsuperscript{-1}K\textsuperscript{-2} after implantation with noble gas ions (Ne, Ar, Xe). The increased power factor is due to a decoupling of the Seebeck coefficient and electrical conductivity identified through a changing scattering mechanism. Ion implantation causes the abundant production of Frenkel pairs, which were found to suppress compensating donors in CuI, and which scenario was also supported by density functional theory calculations. The compensating donor suppression led to a significantly improved Hall carrier concentration, increasing from $6.5\times10^{19}\pm0.1\times10^{19}$\,cm\textsuperscript{-3} to $11.5\times10^{19}\pm0.4\times10^{19}$\,cm\textsuperscript{-3}. This work provides an important step forward in the development of CuI as a transparent conducting material for electronics and thermoelectric generators by introducing beneficial point defects with ion implantation.
\end{abstract}



\maketitle

\noindent Energy harvesting through the thermoelectric effect is a rapidly growing low carbon-emission technology. Materials facilitating such energy conversion are becoming increasingly technologically important \cite{hendricks2022keynote,ruan2024transparent}. Thermoelectric devices convert heat flux into electrical power by the Seebeck effect at a conversion efficiency which depends on the figure of merit $ZT=\alpha^{2}\sigma T/\kappa$ where $\alpha$, $\sigma$, $\kappa$, and $T$ are the Seebeck coefficient, electrical conductivity, total thermal conductivity, and absolute temperature, respectively \cite{zhu2017compromise}. The highest-performing room temperature thermoelectric materials are degenerately doped semiconductors, usually with small band gaps, such as the tetradymite alloys based on (Bi,Sb)\textsubscript{2}(Te,Se)\textsubscript{3} composition, with up to $ZT>1.5$ \cite{pei2020bi2te3}. Unfortunately, these compounds are composed of expensive and toxic precursors \cite{snyder2008complex,zhu2017compromise}, although, recent research has found that non-toxic Mg\textsubscript{3}(Sb,Bi)\textsubscript{2}-based materials can possess up to $ZT=2$ \cite{bano2023realizing,bano2024mg3,wang2024high}. Wide band gap semiconductors provide a wider application scope than their non-transparent counterparts due to their high transparency \cite{liu2021engineering,serhiienko2024record}. Before material integration in devices can be considered on grounds of cost-effectiveness, an increase in the material properties is still required.

The state of the art $n$-type transparent conductors for near room-temperature thermoelectric applications are In\textsubscript{2}O\textsubscript{3}:Sn and ZnO:Al, each with electrical conductivities near $\sigma\sim10000$\,Scm\textsuperscript{-1} \cite{willis2021latest,markwitz2024quasi}, whilst for $p$-type conductors, CuI possesses a moderate conductivity of $\sigma\sim100$\,Scm\textsuperscript{-1} \cite{yang2017transparent,mirza2024cs}. The other $p$-type transparent conductors are doped copper oxides, the best of which is CuAlO\textsubscript{2} with $\sigma=0.01$\,Scm\textsuperscript{-1} \cite{ruttanapun2014reinvestigation}. The reason for the conductivity difference is the high hole mobility in CuI, believed to possibly to reach $30$\,cm\textsuperscript{2}V\textsuperscript{-1}s\textsuperscript{-1} at carrier concentrations of $10^{20}$\,cm\textsuperscript{-3} \cite{willis2023limits}. Presently, hole mobilities are limited to $\approx10$\,cm\textsuperscript{2}V\textsuperscript{-1}s\textsuperscript{-1} at carrier concentrations near $10^{20}$\,cm\textsuperscript{-3}. The high carrier mobilities in-part originate from the band-degenerated light and heavy hole bands with effective masses $0.3m_{0}$ and $2.4m_{0}$, where $m_{0}$ is the free electron mass \cite{grundmann2013cuprous}. The capability for CuI as a functional material has already been demonstrated in laboratory scale components such as in thin film transistor \cite{wu2023high,lee2022synthesis}, optical memory element \cite{mishra2023light,bala2022transparent}, and as $p$-type legs of transparent thermoelectric generator \cite{yang2017transparent,coroa2019highly,morais2018cui} applications, the performance of which in one way or another relies on a high electrical conductivity. Further investigation of its properties and improvements in electrical conductivity toward $1000$\,Scm\textsuperscript{-1} will likely follow with commercial applications in commonplace electronic devices. The further advantages of CuI over alternative $p$-type transparent conductors is its non-toxic and earth-abundant constituents in addition to its facile fabrication procedures \cite{yang2017transparent}. 

Ion implantation is a low-temperature alternative method to annealing which can be used to modify the electrical and thermal properties of materials but is generally limited to application on thin films or the surface layer of a bulk material. Ion implantation has seen recent interest for thermoelectric material thin films, especially the case of ScN \cite{burcea2022influence,burcea2023effect,rao2022multifunctional} or Bi\textsubscript{2}Te\textsubscript{3} \cite{suh2015simultaneous,sinduja2021role} for nanostructuring and chemical doping. Ion implantation is an energetic process which displaces atoms from their lattice sites into interstitial sites and antisites in a collision cascade, leading to the formation of a large concentration of Frenkel pairs within a small volume. The displacement of atoms from their lattice sites usually reduces the phonon thermal conductivity as the long-range crystal order is suppressed, beneficial for thermoelectric materials \cite{tureson2018effect}. Recent studies have investigated the effects of ion implantation of S, Se, and Te as dopants in CuI, but the effect of the implantation damage itself remained unclear \cite{murmu2022role,murmu2024defect,markwitz2024fermi}. Implanting noble gas ions into a material is an excellent testing ground by which to investigate the properties of intrinsic point defects therein. Additionally, it is a process which can be applied without the need for high temperature processing, which generally causes a loss in electrical conductivity of CuI \cite{mulla2018defect,murmu2021effect,almasoudi2022cui,darnige2023insights,storm2020high}.

Noble gas ions provide the opportunity to modify the structure of thin films without also performing chemical modification, allowing the effects to be studied independently. In this work we implanted Ne, Ar, or Xe ions to modify the electronic properties of conducting transparent CuI thin films by introducing point defects. Overall, this process results in an enhanced $p$-type conductivity in CuI. This was identified to be due to an increased Hall carrier concentration with only a minor reduction in Hall carrier mobility. Further investigation on its thermoelectric properties suggest a variation of the scattering process toward ionized impurity scattering. The fast and ambient temperature implementation of this modification process makes it a low cost step during the device fabrication process.

\begin{table}
\caption{\label{tab:SRIMDPA}Relationship between implantation fluence $F$, displacements per ion $D$, and ($70$\,nm) depth-averaged DPA.}
\begin{ruledtabular}
\begin{tabular}{c c c c c}
 Species & Energy [keV] & $F$ [ions\,cm\textsuperscript{-2}] & $D$ [disp.\,ion$^{-1}$] & DPA \\
 \hline
 Ne & $13$ & $0$ & $200$ & $0$ \\ 
 Ne & $13$ & $1.6\times10^{15}$ & $200$ & $1.3$ \\ 
 Ne & $13$ & $3.2\times10^{15}$ & $200$ & $2.5$ \\ 
 Ne & $13$ & $9.6\times10^{15}$ & $200$ & $7.4$ \\ 
 Ne & $13$ & $1.92\times10^{16}$ & $200$ & $14.1$ \\ 
 Ar & $27$ & $0$ & $442$ & $0$ \\ 
 Ar & $27$ & $7.2\times10^{14}$ & $442$ & $1.3$ \\ 
 Ar & $27$ & $1.45\times10^{15}$ & $442$ & $2.5$ \\ 
 Ar & $27$ & $4.34\times10^{15}$ & $442$ & $7.5$ \\ 
 Ar & $27$ & $8.69\times10^{15}$ & $442$ & $14.7$ \\ 
 Xe & $70$ & $0$ & $1331$ & $0$ \\ 
 Xe & $70$ & $5\times10^{14}$ & $1331$ & $2.7$ \\ 
 Xe & $70$ & $1\times10^{15}$ & $1331$ & $5.3$ \\ 
 Xe & $70$ & $2\times10^{15}$ & $1331$ & $10.5$ \\ 
 Xe & $70$ & $3\times10^{15}$ & $1331$ & $15.7$ \\ 
\end{tabular}
\end{ruledtabular}
\end{table}

A previous publication outlines the $60-70$\,nm CuI film deposition by ion beam sputtering \cite{markwitz2023effect}. CuI thin films were deposited on Si(001) substrates for structural and compositional characterization, and soda lime glass substrates for electrical and thermoelectric property characterization. After deposition the CuI films were implanted with neon, argon, or xenon, at $13$\,keV, $27$\,keV, or $70$\,keV, respectively, to provide equivalent projected ranges ($26$\,nm) and straggles ($15$\,nm), with the implantation fluences scaled to approximately match the displacements per atom (DPA) calculated with the software Stopping Range of Ions in Matter (SRIM) in the \textit{Detailed Calculation with full Damage Cascades} mode \cite{ziegler1985stopping}. Due to the strong texture of the CuI films, the implantation was conducted at $7$\,{\degree} from the sample normal. The relationship between implantation fluence ($F$), the displacements per ion ($D$), the atomic density ($\rho_{n}=3.385\times10^{22}$\,at\,cm\textsuperscript{-3}), and the resultant DPA (averaged over the $70$\,nm initial film thickness) are written as $\text{DPA}=FD/\rho_{n}$, and summarized in Table \ref{tab:SRIMDPA}. The samples were not visibly affected by implantation. Throughout this work the samples are labelled by the implantation species with the associated DPA (see Table \textcolor{red}{S1} for a summary of sample details). The implantation and DPA depth profiles are included as Figures \textcolor{red}{S1(a-f)}. The implantation energies were chosen to just avoid the effects of film-substrate mixing, which comes at a trade-off with an imperfect defect uniformity throughout the depth of the films, providing the opportunity to investigate the effects of the added defects throughout the film. More details pertaining to the implantation method are covered in previous publications \cite{markwitz2024fermi,murmu2024defect}.

\begin{figure}
    \centering
    \includegraphics[width=\columnwidth]{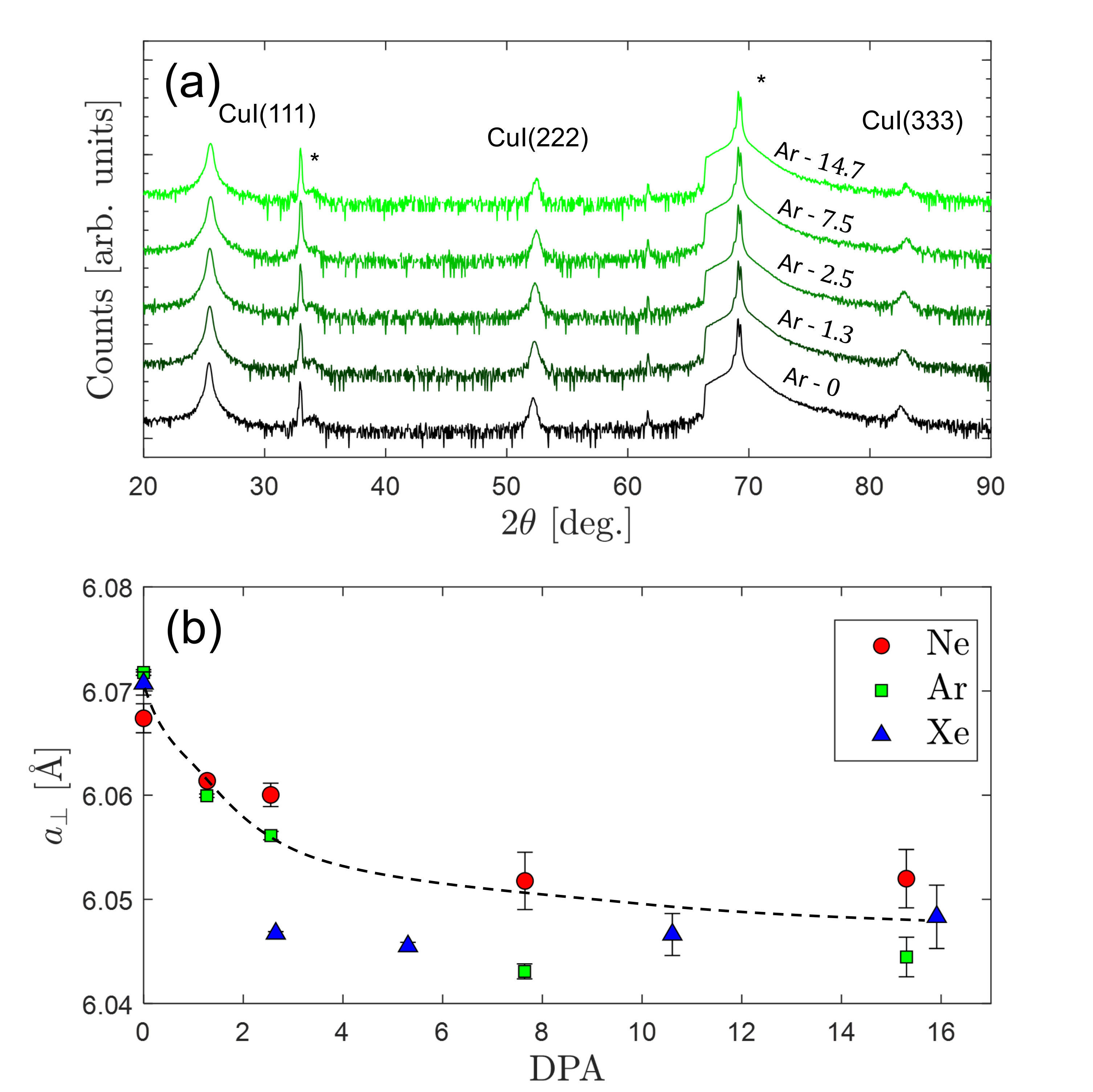}
    \caption{(a) Angle-symmetric X-ray diffraction patterns, vertically offset for visual clarity. (b) Out-of-plane lattice constants derived from the XRD measurement, the dashed line is used as a guide.}
    \label{fig:characterization}
\end{figure}

\begin{figure*}
    \centering
    \includegraphics[width=\textwidth]{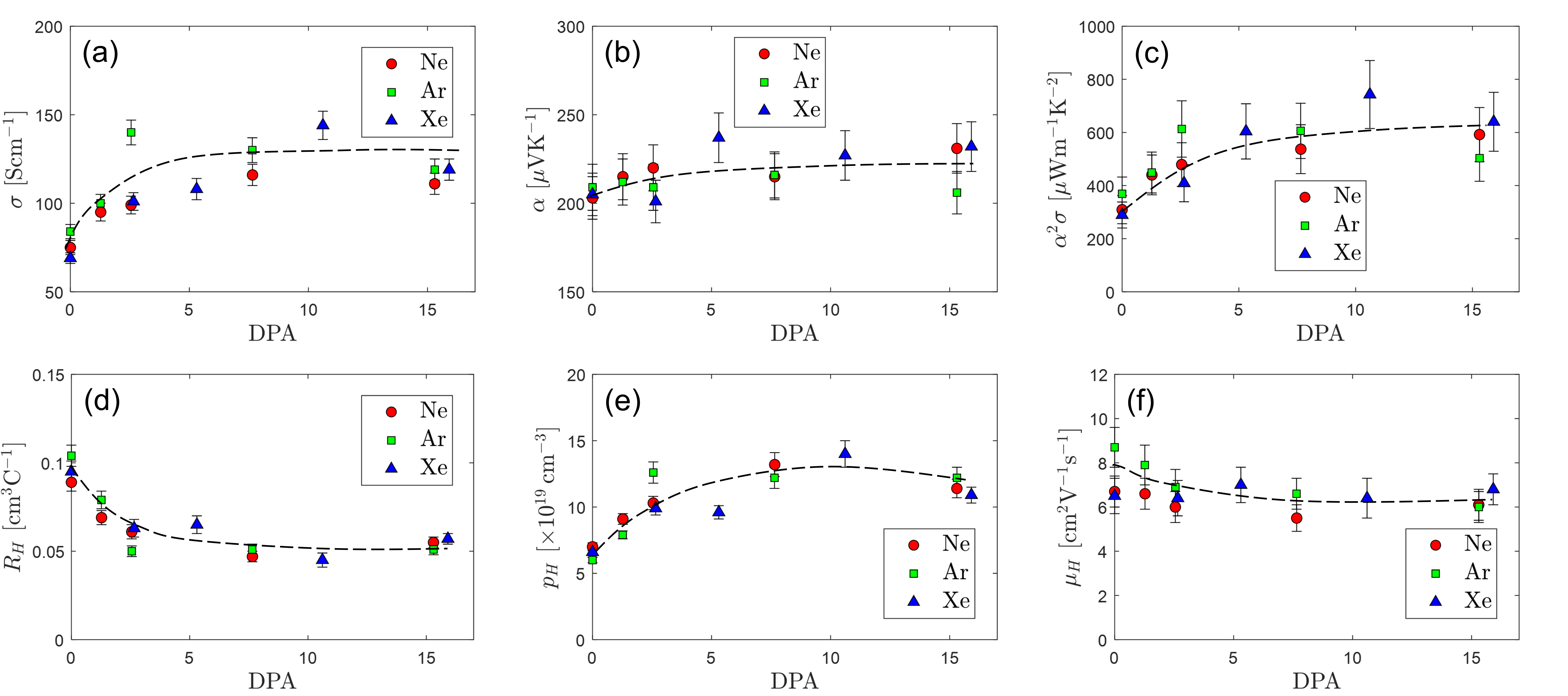}
    \caption{(a) Electrical conductivity, (b) Seebeck coefficient, (c) Hall coefficient, (d) power factor, (e) Hall carrier concentration, and (f) Hall carrier mobility of CuI films implanted with noble gas ions. The dashed lines are used as a guide.}
    \label{fig:transportMeasurements}
\end{figure*}

The films' composition and thicknesses were investigated with Rutherford backscattering spectrometry (RBS). The RBS measurements were conducted with a $2.0$\,MeV \textsuperscript{4}He\textsuperscript{+} beam with a $165$\,{\degree} backscattering angle, a current density of $10$\,nA and an integrated charge of $20$\,{\textmu}C with a surface barrier detector \cite{kennedy2007ion}. The backscattering spectra are shown in Figures \textcolor{red}{S2(a-c)}, which suggest that the as-deposited films possessed stoichiometric [Cu]/[I] ratios of $1.02\pm0.02$. The implanted samples exhibited increasing [Cu]/[I] ratios to an average of $1.16\pm0.07$ when implanted to an average DPA of $15$, the ratios of which are included in Figure \textcolor{red}{S3}. Excess copper in CuI is commonly observed for thin films \cite{storm2020high,darnige2023insights,zhu2011transparent,bae2023precision}. The loss of iodine, and excess of copper is a result of preferential halide sputtering, subsequently oxidizing the film surface upon exposure to air \cite{townsend1987optical,storm2021evidence,crovetto2020water}. The Ne in the Ne-implanted films could not be measured with RBS (Figure \textcolor{red}{S2a}) due to the measurement noise provided by the silicon substrate. The Ar in the Ar-implanted films was observed with RBS (Figure \textcolor{red}{S2b}). The presence of Xe in the Xe-implanted films by RBS (Figure \textcolor{red}{S2c}) is inconclusive due to the similar atomic mass of Xe and I.


The films' structural properties were studied with X-ray diffraction using a Rigaku SmartLabs diffractometer, employing a Cu X-ray source. The angle-symmetric measurements are shown for the argon implanted samples in Figure \ref{fig:characterization}\textcolor{red}{a} and Figures \textcolor{red}{S4(a-b)} for the others, resulting in the identification of the zincblende CuI, strongly textured along the $\langle$111$\rangle$ direction \cite{grundmann2013cuprous}. The lattice constants of the as-deposited films are $6.070\pm0.002$\,{\AA}, settling to $6.048\pm0.004$\,{\AA} when implanted to an average DPA of $15$, calculated using the goniometer error function \cite{storm2020high}. These are shown in Figure \ref{fig:characterization}\textcolor{red}{b}, and the fits themselves depicted in Figures \textcolor{red}{S5(a-c)}. The out-of-plane lattice constant of CuI thin films are well known to be greater than the bulk value of $6.054$\,{\AA}, but is known to not be strongly dependant on the [Cu]/[I] ratio \cite{grundmann2013cuprous,storm2020high}. CuI thin films grown along the (111) plane is known to be rhombohedrally distorted, associated with a contraction (an expansion) of the out-of-plane (in-plane) lattice constants \cite{nakamura2022band}. 

\begin{figure*}
    \centering
    \includegraphics[width=\textwidth]{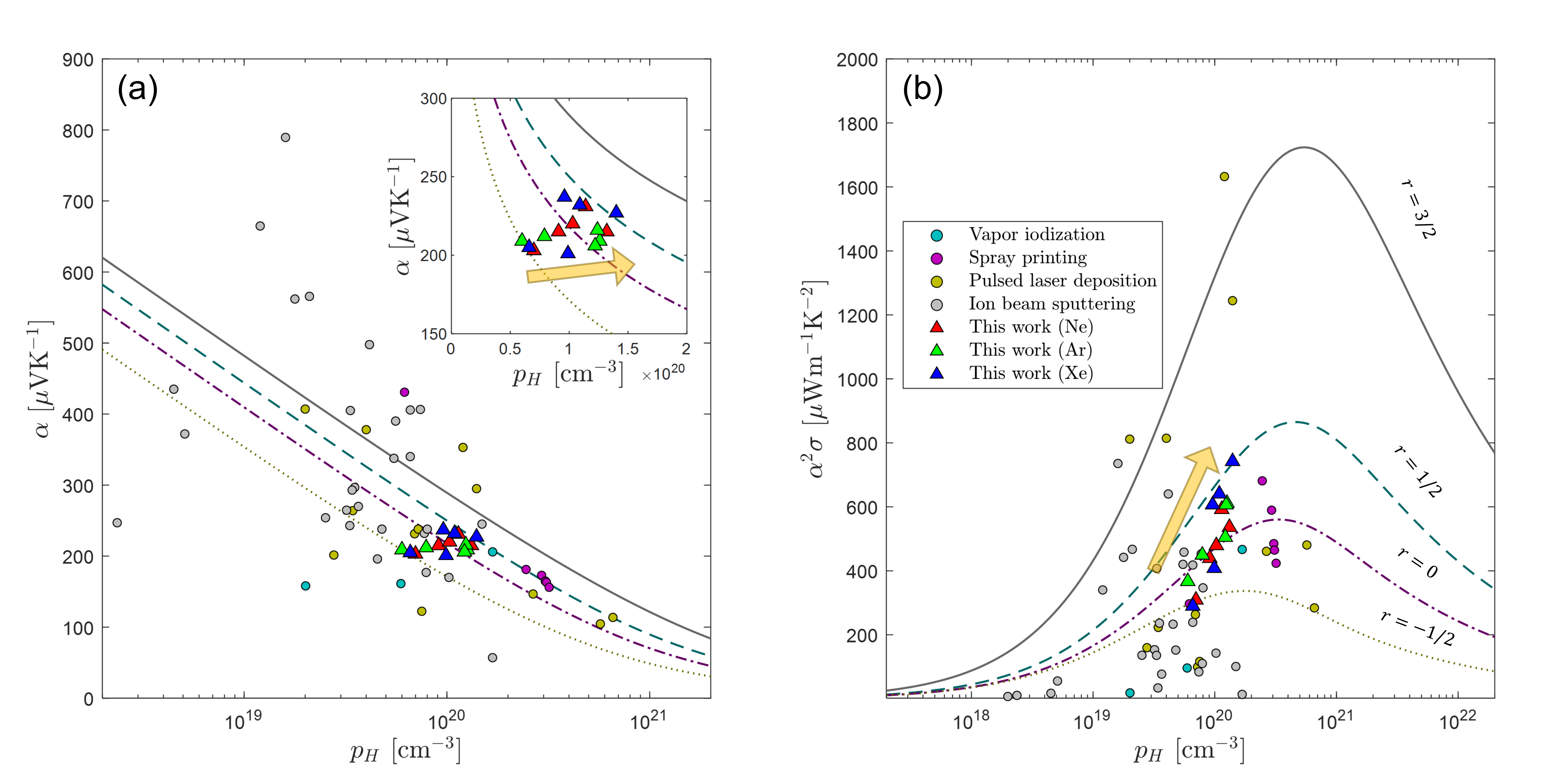}
    \caption{(a) Seebeck coefficient and Hall carrier concentration plot with the corresponding Boltzmann transport equation solutions with a variety of energy-dependencies of the scattering time ($r=-1/2$ to $r=3/2$). A survey of literature data \cite{almasoudi2022cui,bae2023precision,mirza2023role,coroa2019highly,murmu2021effect,murmu2022role,murmu2024defect,markwitz2024fermi} (colored circles) is compared to the experimental findings of neon (red, squares), argon (green, diamonds), and xenon (blue, triangles). Inset graph presents a magnified view of the observed Hall carrier concentration and Seebeck coefficients. (b) Power factor plotted against Hall carrier concentration for the same data from literature. Arrows are used to imply the progression of the thermoelectric material properties with increasing DPA.}
    \label{fig:Literature}
\end{figure*}

To evaluate the thermoelectric and carrier properties of the CuI thin films, the Seebeck coefficient $\alpha$, electrical conductivity $\sigma$, and Hall coefficient $R_{H}$ were measured. Room temperature Seebeck coefficient measurements were conducted with an ADVANCE RIKO ULVAC ZEM-3 with pressed nickel contacts. Hall effect measurements were conducted with an HMS-3000 after sputtering gold contacts on the corners of the samples. The systematic measurement errors for the Seebeck effect measurements are $6$\,\% due to systematic uncertainties, and $5$\,\% for and Hall effect measurements \cite{alleno2015invited}. The results of those measurements are summarized in Figure \ref{fig:transportMeasurements}\textcolor{red}{(a-f)}. The Seebeck and Hall coefficients were positive for all samples. The electrical conductivity increased from $76\pm2$\,Scm\textsuperscript{-1} to $116\pm3$\,Scm\textsuperscript{-1} for the unimplanted, and samples with highest DPA, respectively, an increase of $53\pm6$\,\%. Qualitatively, the change in electrical conductivity saturates for a DPA of $5$ for all implantation species, similar to the results of Burcea \textit{et al.} \cite{burcea2023effect} for $n$-type ScN with the direct-impact model as proposed by Gibbons \cite{gibbons1972ion}. The Seebeck coefficient slightly increased between the unimplanted samples ($206\pm7$\,{\textmu}VK\textsuperscript{-1}) and the samples with highest DPA ($223\pm8$\,{\textmu}VK\textsuperscript{-1}). Notably, the usual relationship between the electrical conductivity and Seebeck coefficient was overcome by noble gas ion implantation. The increasing electrical conductivity results in the improvement of the power factor, calculated from $\alpha^{2}\sigma$, the values ranging from $322\pm32$\,{\textmu}Wm\textsuperscript{-1}K\textsuperscript{-2} to $578\pm58$\,{\textmu}Wm\textsuperscript{-1}K\textsuperscript{-2}. Overall, the highest power factor was observed for the Xe-10.5 sample with a power factor of $743\pm128$\,{\textmu}Wm\textsuperscript{-1}K\textsuperscript{-2}. This result exceeds the highest CuI thin film power factors of Coroa \textit{et al.} \cite{coroa2019highly} ($467$\,{\textmu}Wm\textsuperscript{-1}K\textsuperscript{-2}), Bae \textit{et al.} \cite{bae2023precision} ($681$\,{\textmu}Wm\textsuperscript{-1}K\textsuperscript{-2}), and Mirza \textit{et al.} \cite{mirza2023role} ($481$\,{\textmu}Wm\textsuperscript{-1}K\textsuperscript{-2}), but lower than those of Almasoudi \textit{et al.} \cite{almasoudi2022cui} ($1632$\,{\textmu}Wm\textsuperscript{-1}K\textsuperscript{-2}). Concomitant with the increase in electrical conductivity, there was a reduction in the average Hall coefficient from $0.096\pm0.010$\,cm\textsuperscript{3}C\textsuperscript{-1} to $0.054\pm0.005$\,cm\textsuperscript{3}C\textsuperscript{-1}. The Hall coefficient is related to the carrier concentration in the single parabolic band model by $p=\left(qR_{H}\right)^{-1}$, wherein $q$ is the elementary charge. The derived Hall carrier concentrations increase from an average of $6.5\times10^{19}\pm0.1\times10^{19}$\,cm\textsuperscript{-3} to an average of $11.5\times10^{19}\pm0.4\times10^{19}$\,cm\textsuperscript{-3}, an increase of $77\pm4$\,\%. Also, the Hall carrier mobility can be derived from $\mu_{H}=R_{H}\sigma$, reducing slightly from an initial value of $7.3\pm0.5$\,cm\textsuperscript{2}\,V\textsuperscript{-1}s\textsuperscript{-1} to $6.3\pm0.4$\,cm\textsuperscript{2}\,V\textsuperscript{-1}s\textsuperscript{-1}. Such an effect, where the electrical conductivity increases by ion irradiation with a $-16\pm14$\,\% reduction in Hall mobility has before been noted in Bi\textsubscript{2}Te\textsubscript{3} which resulted in an improved thermoelectric power \cite{suh2015simultaneous}. It is possible that the point neutral and ionized disorder introduced by implantation is the cause for the reduction in carrier mobility. Willis \textit{et al.} \cite{willis2023limits} summarize the state-of-the-art to which the results of this work can be directly compared.

The combined variation in Hall carrier concentration, Seebeck coefficient, and Hall carrier suggest that there is a change in the carrier scattering process. To further investigate this, the semiclassical Boltzmann transport model in the relaxation time approximation is applied \cite{markwitz2023effect}. To investigate variations in scattering mechanism the Seebeck coefficient is plotted against the Hall carrier concentration in Figure \ref{fig:Literature}\textcolor{red}{a}, in addition to the corresponding theoretical curves. The results are also compared to results from CuI thin films in literature \cite{almasoudi2022cui,bae2023precision,mirza2023role,coroa2019highly,murmu2021effect,murmu2022role,murmu2024defect,markwitz2024fermi}. The inset Figure \ref{fig:Literature}\textcolor{red}{a} presents a magnified view of the results, which suggests a deviation from the conventional relationship of decreasing $\alpha$ with increasing $p_{H}$. Such a decoupling effect be attributed to a change of scattering mechanism toward ionized impurity scattering similar to the effect observed by Suh \textit{et al.} \cite{suh2015simultaneous} in Bi\textsubscript{2}Te\textsubscript{3}. The power factor is compared to the Hall carrier concentration and the corresponding Boltzmann transport equations in Figure \ref{fig:Literature}\textcolor{red}{b} (by using $\tau_{0}=6.5$\,fs). The power law relaxation time was used for this calculation, with $r$ the energy dependence of the scattering time, varying from $r=-1/2$ for acoustic phonon scattering, $r=0$ for neutral impurity scattering, $r=1/2$ for polar optical phonon scattering, and $r=3/2$ for ionized impurity scattering. The scattering process is implied to change as the data does not track along any particular $r$ curve, instead traversing towards the $r=3/2$ curve. The highest-performing samples across literature possess Hall carrier concentrations of the order of $10^{20}$\,cm\textsuperscript{-3}.

\begin{figure*}
    \centering
    \includegraphics[width=\textwidth]{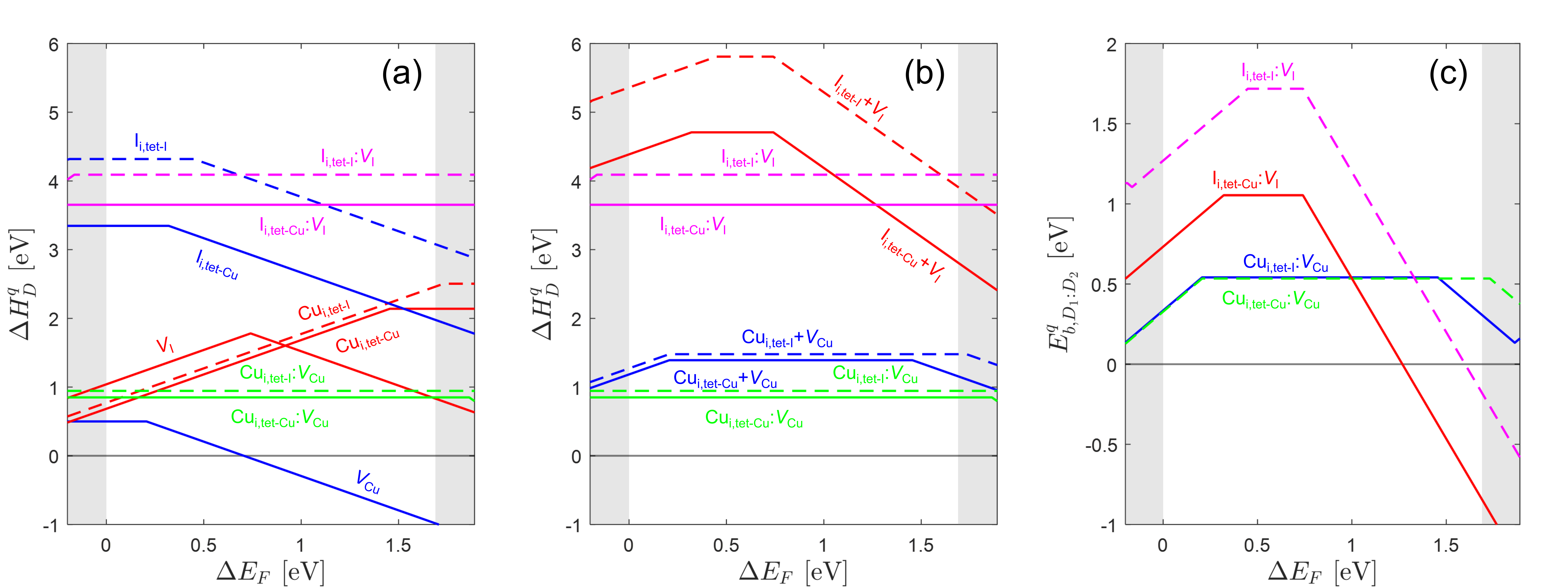}
    \caption{(a) Formation energy of isolated and Frenkel defects. (b) Summed formation energies of the relevant isolated defects compared to their defect complexes. (c) Frenkel pair binding energy diagram. The defects involving the interstitial atoms coordinated by iodine (copper) atoms are drawn as dashed (filled) lines.}
    \label{fig:DFT}
\end{figure*}

The cause for the increased carrier concentration is presumably due to the increased concentration of copper vacancies, or alternatively, an increased concentration of interstitial iodine, although those possess a greater formation energy \cite{huang2012first}. It is also possible that more complicated defect complexes could be formed as a result of the implantation which act as shallow acceptor states that could also increase the observed Hall carrier concentration \cite{burcea2023effect}. Darnige \textit{et al.} \cite{darnige2023insights} hypothesized that during thin film growth, a large concentration of Frenkel pairs (complexes of vacancies and interstitials of the same type) are included which can be annealed out, resulting in a reduced carrier concentration after annealing \cite{murmu2021effect,almasoudi2022cui,mulla2018defect}. In addition to those produced during growth, Frenkel pairs are produced in abundance through ion implantation, especially in ionic crystals, which could be the reason for the increased carrier concentration \cite{townsend1987optical}. A connection between Frenkel pairs and the macroscropic electrical properties can be made by use of density functional theory (DFT) calculations, such as was done for Cd\textsubscript{1-x}Zn\textsubscript{x}Te \cite{jakubas2008generation}, CeO\textsubscript{2} \cite{smith2023structural}, ThO\textsubscript{2} \cite{moxon2022structural}, and CsPb(I\textsubscript{1-x}Br\textsubscript{x})\textsubscript{3} \cite{sabino2023light}.

We conduct density functional theory calculations for intrinsic defects in CuI, such as vacancies (\textit{V}\textsubscript{Cu}/\textit{V}\textsubscript{I}), antisites (Cu\textsubscript{I}/I\textsubscript{Cu}), and interstitials (Cu\textsubscript{i, tet-I}/I\textsubscript{i, tet-I}/Cu\textsubscript{i, tet-Cu}/I\textsubscript{i, tet-Cu}), using the same computational setup using the PBE exchange-correlation functional as discussed in our previous work \cite{murmu2024defect,markwitz2024fermi}. The interstitial sites are coordinated tetragonally either by Cu ions, or I ions \cite{grauvzinyte2019computational}. The calculated formation and thermodynamic transition energies are in good agreement with the work of Huang \textit{et al.} \cite{huang2012first}, for both the Cu-rich and Cu-poor chemical potential limits. Taking a step further, the Frenkel pair binding energy is calculated with

\begin{equation}
    E_{b,D_{1}:D_{2}}^{q}=\Delta H_{D_{1}}^{q}+\Delta H_{D_{2}}^{q}-\Delta H_{D_{1}:D_{2}}^{q}
    \label{eq:bindingenergy}
\end{equation}

\noindent which relates the formation energy of the isolated defects ($\Delta H_{D_{1}}^{q}$ and $\Delta H_{D_{2}}^{q}$) with the formation energy of the defect complex ($\Delta H_{D_{1}:D_{2}}^{q}$) \cite{jakubas2008generation,ananchuensook2024hydrogen}. A positive binding energy ($E_{D_{1}:D_{2}}^{q}$) indicates that the defects are stable complexes and will remain in proximity of one another, but, it does not indicate the likelihood of the formation thereof, which relies on a low formation energy of the defect complex ($\Delta H_{D_{1}:D_{2}}^{q}$). Figure \ref{fig:DFT}\textcolor{red}{a} shows the formation energy of the isolated defects and and defect complexes in CuI, Figure \ref{fig:DFT}\textcolor{red}{b} shows the summed isolated defect formation energies, while Figure \ref{fig:DFT}\textcolor{red}{c} shows the Frenkel pair defect binding energies calculated using Eq. \ref{eq:bindingenergy}.

Due to the low formation energy of \textit{V}\textsubscript{Cu} the self-consistently calculated Fermi energy is always near the valence band edge \cite{grauvzinyte2019computational}. When the Fermi energy is near the valence band edge, the pair of isolated defects (\textit{V}\textsubscript{Cu} and Cu\textsubscript{i, tet-I}/Cu\textsubscript{i, tet-Cu}) overall exhibit donor status. On the other hand, the charge of the Frenkel pairs remains neutral, which acts to passivate the compensating donors. Such a phenomenon has been previously used to describe the relation to macroscopic conductivity-switching effect based on different charge states of the isolated defects and their Frenkel pair binding energies in Cd\textsubscript{1-x}Zn\textsubscript{x}Te \cite{jakubas2008generation}. Additionally, the effect of hydrogen passivation of cation vacancies in CuMO\textsubscript{2} (M=Al, Ga, In) was similarly investigated \cite{ananchuensook2024hydrogen}. This donor passivation effect is exacerbated by ion implantation, where Frenkel pairs are produced in abundance within the CuI matrix, leading to an overall $p$-type doping effect.

In summary, noble gas ion implantation with Ne, Ar, or Xe is a post-deposition technique to modify the transparent conducting and thermoelectric properties of CuI thin films by introducing intrinsic point defects. The out-of-plane lattice constant reduced from the as-deposited values of $6.070\pm0.002$\,{\AA} to $6.048\pm0.004$\,{\AA} for the highest-implanted films. Simultaneously, there was a remarkable improvement in electrical conductivity from an average of $76\pm2$\,Scm\textsuperscript{-1} to $116\pm3$\,Scm\textsuperscript{-1} by implantation to an average DPA of $15$, regardless of the noble gas ion. This was attributed to the increase in Hall carrier concentration, which increased from $6.5\times10^{19}\pm0.1\times10^{19}$\,cm\textsuperscript{-3} to $11.5\times10^{19}\pm0.4\times10^{19}$\,cm\textsuperscript{-3}. The increasing Hall carrier concentration could be due to (1) the formation of Frenkel pairs which suppress the concentration of compensating donors, or (2) the replenishment of O into \textit{V}\textsubscript{I} near the film surface. It should be noted that O is known to not be a very shallow acceptor within CuI with an acceptor ionization energy between $0.145$\,eV and $0.28$\,eV, and is therefore not expected to be the reason why the carrier concentrations are driven above $10^{20}$\,cm\textsuperscript{-3} \cite{bar2024deconvolution,grauvzinyte2019computational,huang2012first}. 

The power factor of CuI thin films increased from an average of $322\pm32$\,{\textmu}Wm\textsuperscript{-1}K\textsuperscript{-2} to $578\pm58$\,{\textmu}Wm\textsuperscript{-1}K\textsuperscript{-2} by ion implantation with noble gas ions, comparable to the state-of-the-art power factors reported in literature. This improvement in the thermoelectric properties is likely due to a change in scattering mechanism from phonon to ionized impurity scattering, achieved by ambient temperature defect engineering with ion implantation. Further research should investigate the effects of the order of implantation and annealing on the properties and stability of the microstructural and transport properties of CuI.




\begin{acknowledgments}
This research is funded the Royal Society of New Zealand through the Marsden Fund, New Zealand (grant number MFP-GNS2301), the Ministry of Business, Innovation and Employment and the JST Mirai Program, Japan (grant number JPMJMI19A1). We acknowledge Mr. Chris Purcell and Mr. Niall Malone (National Isotope Centre, GNS Science, New Zealand) for performing the RBS measurements and for helpful discussion.
\end{acknowledgments}

\section*{Supplementary Material}
See the supplementary material for more details on the simulation of the noble gas ion implantation and DPA depth profiles, the measured Rutherford backscattering spectra and X-ray diffraction patterns, and derived quantities from those data.

\section*{References}

\providecommand{\noopsort}[1]{}\providecommand{\singleletter}[1]{#1}%

\end{document}